\documentclass[final,3p,times,twocolumn]{elsarticle}
\usepackage{amsmath,amssymb}
\usepackage{graphicx}
\journal{Physics Letters B}

\begin{document}
\begin{frontmatter}
\title{ Diffraction-corrected neutrino flux and  $\nu N$ total cross section }
\author{Kenzo Ishikawa and Yutaka Tobita}
\address{Department of Physics, Faculty of Science, Hokkaido
University Sapporo 060-0810, Japan}
\begin{abstract}
A neutrino produced in a particle decay maintains an unusual  wave nature 
that  reveals a diffraction phenomenon in a large spatial region. 
Diffraction gives a new component to a neutrino  flux 
in this  region and  modifies  determinations   
of physical quantities in neutrino experiments. A total cross
 section of neutrino
 nucleon scattering at  high energies  is studied in this paper.  
 A cross section is proportional to an energy in 
the quark-parton model and is modified with 
the energy-dependent neutrino flux induced by 
the diffraction component. A total number of events  devided by the
 energy is not constant but has a slight energy dependence.  
 Theoretical values agree  well with the recent experiments.  
\end{abstract}

\end{frontmatter}
\section{Neutrino diffraction and absolute neutrino  flux} 

In high energy neutrino experiments, neutrinos from pion decays are  used. Neutrino flux is not directly
measurable and 
is estimated  from a pion's flux and decay probabilities. 
 Now  precision measurements are becoming possible in neutrino 
experiments and a precise value of the flux is required. So  it is
important   to verify  a standard assumption and to find a precise
value.  We point out that a correction due to a quantum mechanical
interference of a single  neutrino  is necessary in 
the flux in near detector regions.

Decay amplitudes and probabilities  are obtained normally with  a standard
method of S-matrix using  plane waves. 
An  asymptotic boundary condition \cite{LSZ,Low} is assumed and an
average decay rate 
is found and used for estimations of the flux. This 
method  gives precise values of fluxes of  final states  in the 
asymptotic space-time region  where a particle  has no correlation and
is described with a plane wave.  It has 
never been clarified, however, if  particles in the final state are 
observed in an asymptotic region.  It would be reasonable to assume that 
a boundary between an asymptotic and non-asymptotic regions is of 
microscopic size, since transitions  of weak decays occur in short distance
regions. However we find that is not the case. An asymptotic region  for 
a neutrino in a particle decay is located in a
macroscopically distant area from an initial particle's position.  Waves
produced at different positions accumulate at a detector and make experiments
in the inside of this area unusual. Superposition of waves  causes
 an interference effect in a form of a diffraction in a parallel  
direction to the momentum, which  makes the plain waves 
non-asymptotic.  Various relations satisfied for the
asymptotic states, such as a unitarity, an energy momentum conservation,
and others are modified with the diffraction term.       

To study the interference phenomenon of particles produced in a decay,
it is convenient to study a 
coherence length from a behavior of a correlation function defined
with a  wave function. Let $\psi(x,\gamma)$ be a wave function of
a final state composed of one particular particle at a space-time
position $x$ and others, $\gamma$. An amplitude of observing this
particle at a  momentum $p$  and  a position $X$  using a detector 
is defined with  the 
wave function  at one space-time position $x$ and a wave packet
$w(p,x)$ in the form
\begin{eqnarray}
\Psi(X,p,x,\gamma)=w(p,x-X) \psi(x,\gamma),
\end{eqnarray}
where  $w(p,x-X)$ is a wave packet with which this 
particle interacts and a Gaussian form with a  size
$\sigma$ and velocity ${\vec v}={\vec p}/E,E=\sqrt{\vec{p}^{\,2}+m^2}$
is used for the sake of simplicity, 
\begin{align}
w(p,x-X)=N_0 e^{-\frac{\left({\vec x}-{\vec X}-{\vec
 v}(x^0-X^0)\right)^2}{2\sigma}-ip\cdot(x-X)},
\end{align}
where $p\!\cdot\!x$ is scalar product of a four dimensional coordinate $x$ and
momentum $p$ and $N_0$ is a normalization factor. 

Since the wave packet vanishes at $|{\vec x}\,|\rightarrow \infty $, it 
satisfies the asymptotic boundary condition. So the wave packet is
approapriate to study a finite size correction. An integral of the 
product $\Psi^{*}(X,p,x_1,\gamma)\Psi(X,p,x_2,\gamma)$
is the probability $C(X,p)$,
\begin{align}
C(X,p)= \int   d^4x_1 d \gamma d^4x_2  \Psi(X,p;,x_1,\gamma)^{*}
 \Psi(X,p;x_2,\gamma),\nonumber
\end{align}
where $d\gamma$ stands for the volume element of state $\gamma$. $C(X,p)$   
is written as
\begin{eqnarray}
 \int d^4 x_1 d^4x_2  w^{*}(p,x_1-X)w(p,x_2-X)  \Delta(x_1,x_2),\nonumber
\end{eqnarray}
using the  correlation function defined with
\begin{eqnarray}
\Delta(x_1,x_2)=\int d \gamma \psi(x_1,\gamma)^{*} \psi(x_2,\gamma).
\end{eqnarray}
The correlation function generally has the  singular function  
\begin{eqnarray}
\delta(\lambda)\epsilon(x_1^0-x_2^0),\lambda=(x_1-x_2)^2,
\end{eqnarray}
or regular functions $e^{i\tilde m\lambda}, e^{-\tilde m
 \lambda}$ 
where a finite  $\tilde m$ is determined from 
dynamics of the system. If one  momentum  that is conjugate 
to $x_1-x_2$ is integrated and the singular function $\delta(\lambda)$ is 
derived, this gives the form 
\begin{eqnarray}
C(X,p) =C^{(1)}e^{-\frac{|{\vec X}|}{l_0}},\   |\vec{X}| > l_0,
\end{eqnarray}
where  $l_0$ is given in the form 
\begin{eqnarray}
l_0=\frac{2|\vec{p}\,| \hbar c}{m^2},
\end{eqnarray}
in a high energy region, $|\vec{p}\,| \gg m$.
Other oscillating  or decreasing  terms  give constants  to
$C(X)$. So $l_0$ stands for a coherence length of
the wave function. 
The coherence lengths of  a pion
and an electron of an energy of $1$ [GeV] are 
\begin{eqnarray}
& &l_0^{pion}={2 \hbar c \over 0.13^2}\,[\text{GeV}^{-1}]= 2\times 10^{-14}\, [\text{m}],\\
& &l_0^{electron}={2 \hbar c \over 0.5^2}\,[\text{GeV}^{-1}]\approx  10^{-10}\, [\text{m}],\label{electron-coherence}
\end{eqnarray} 
and those of other hadrons are shorter than that of a pion. So charged 
leptons and hadrons have short coherence lengths of  microscopic
sizes in an ordinary high energy region. 
Neutrinos are exceptional and lighter
than an electron by $10^6$ or more and have coherence lengths,
\begin{eqnarray}
l_0^{neutrino}\approx 10^2-10^3\,[\text{m}], 
\label{neutrino-coherence}
\end{eqnarray} 
which is a macroscopic size.
From Eqs.\,$(\ref{electron-coherence})$ and  $
(\ref{neutrino-coherence})$, the non-asymptotic 
region is narrow in charged leptons and very wide in neutrinos.

In  outside of the coherence  length $|{\vec X}| \gg l_0$, the 
wave-like correlation  vanishes. Hence the state  
behaves like  a particle-like object which has no
spatial correlation and  satisfies the asymptotic condition.   
In  inside, $|{\vec X}|<l_0$, on the other hand, the
probability varies with $|{\vec X}|$ and a wave-like correlation remains.  
So the physical state   behaves like an unusual wave that is distinct from 
a particle-like object .  The physical state is not 
treated as a real particle and does not satisfy the asymptotic condition.   
So the coherence 
length is used  as the  boundary  between the asymptotic and
non-asymptotic regions. 

Transition probabilities  in the asymptotic region are studied using an 
ordinary S-matrix and a decay amplitude of plane waves and hold
relations of a standard S-matrix. But  those 
in the non-asymptotic region are  peculiar and  different from those of
particle natures of a standard S-matrix. They   violate some relations
and   are  not studied with a standard S-matrix of plane waves.  The 
physics in both regions can be  studied with a time  dependent 
Schr\"{o}edinger equation  or a scattering amplitude of wave packets,
which satisfy the asymptotic 
conditions and have manifest position dependence. 
We study the physics in the non-asymptotic region using   the
position dependent amplitudes.

Charged leptons are massive and their measurements are made in 
their asymptotic regions. So physical quantities are computed
with the ordinary S-matrix and agree with  values obtained in experiments. 
 Weak decays of  mesons are understood well with $V-A$ weak 
interactions through measurements of the charged leptons. 
For  neutrinos the situation is different. Since the asymptotic region is
located in a distant area, neutrinos may be  observed   in 
non-asymptotic regions. Then the neutrino  reveals wave-like phenomena   
such as interference or diffraction and has an unusual position dependence.
A diffraction term, in fact,  emerges and gives a finite contribution to
 the neutrino flux. A new term which does not satisfy varios relations
 of the standard S-matrix is added to  an amplitude in a short distance 
region.   The total
neutrino's flux thus obtained deviates from that obtained using the naive decay
probability and has an excess.   Implications of the diffraction term to 
the total cross sections \cite{particle-data,excess-near-detectorNOMAD,excess-near-detectorMino}
are studied.

\section{Position dependent probability}  
A
diffraction term   \cite{Ishikawa-Tobita-prl1,Ishikawa-Tobita-prl2} is derived   from a position-dependent amplitude 
of a pion decay process. 
The
amplitude   is determined in the first order of the $V-A$ weak
Hamiltonian $H_{w}$ in the form, 
$T=\int d^4x \, \langle l,{\nu}
 |H_{w}(x)| \pi \rangle$, here  a pion is 
prepared at a time  $\text{T}_\pi$,  and  a neutrino is measured with  a 
wave packet of 
 a space-time position $(\text{T}_{\nu},
\vec{\text{X}}_{\nu})$ and a charged lepton is un-measured.  
 The neutrino wave packet
 \cite{Ishikawa-Shimomura,Ishikawa-Tobita-ptp,Ishikawa-Tobita} expresses
  a target nucleon which the neutrino interacts with. Hence the
 wave packet is well localized in the coordinate variable. To represent
 this property, the momentum variables must cover whole momentum
 region and the Gaussian form is used in this work.        
A set of wave packets of a $\sigma_{\nu}$ with  a continius spectrum of 
momentum  and coordinate  is complete \cite{Ishikawa-Shimomura}. Hence
neutrinos are  described 
with  wave packets of  central values of the momentum and  coordinate
and a width \cite{Kayser,Giunti,Nussinov,Kiers,Stodolsky,Lipkin,Asahara}.
Other particles  are described with  the plane waves.  They are
expressed in the form   
$
|\pi \rangle=   | {\vec p}_{\pi},\text{T}_{\pi}  \rangle,\ 
|l ,\nu \rangle=   |l,{\vec p}_l ;\nu,{\vec p}_{\nu},\vec{\text{X}}_{\nu},\text{T}_{\nu}          \rangle.$
The
amplitude $T$ is
written with the hadronic $V-A$ current and  Dirac spinors  in the form
\begin{align}
\label{amplitude}
T &= \int d^4xd{\vec k}_{\nu}
\,N_1\langle 0 |J_{V-A}^{\mu}(x)|\pi \rangle 
\nonumber\\
&\times \bar{u}({\vec p}_l)\gamma_{\mu} (1 - \gamma_5)\nu({\vec k}_{\nu})e^{ip_l\cdot x + 
ik_\nu\cdot(x - \text{X}_\nu)
 -\frac{\sigma_{\nu}}{2}({\vec k}_{\nu}-{\vec p}_{\nu})^2},  
\end{align}
where 
$N_1=ig \left({\sigma_\nu/\pi}\right)^{\frac{4}{3}}\left({m_l m_{\nu}}/{
 E_l E_{\nu}}\right)^{\frac{1}{2}}$, and  the time $t$ is
 integrated in the region $\text{T}_{\pi} \leq t$. 
${\sigma_{\nu}}$ is  
a neutrino wave packet size and is estimated   using  a nucleus size.
The Gaussian form of the wave packet is used  for a sake
of simplicity but the most important result is unchanged 
in general wave packets as was verified in 
\cite{Ishikawa-Tobita-prl1}. 

If the coordinate ${\vec x}$ is integrated in Eq.\,$(\ref{amplitude})$,
the delta function $(2\pi)^3\delta({\vec p}_{\pi}-{\vec p_l}-{\vec k}_{\nu})$ emerges. So ${\vec k}_{\nu}$ is integrated easily and the
Gaussian part becomes $e^{-\frac{{\sigma}_{\nu}}{2}({\vec p}_{\pi}-{\vec
p}_l-{\vec p}_{\nu})^2+iE({\vec p}_{\pi}-{\vec p}_l-{\vec p}_{\nu})(x^0-T_{\nu})-i({\vec p}_{\pi}-{\vec
p}_l-{\vec p}_{\nu})\cdot\vec{\text{X}}_{\nu}}$.  The exponent  has two
stationary momenta, ${\vec p}_l \approx{\vec p}_{\pi}-{\vec p}_{\nu}$
and  ${\vec v}(x^0-\text{T}_{\nu})-\vec{\text{X}}_{\nu} \approx 0,{\vec v}={\partial
E({\vec p}_{\pi}-{\vec p}_l-{\vec p}_{\nu}) \over \partial {\vec p_l}}$ \cite{Ishikawa-Shimomura}.  
Hence the probability gets  a
contribution from a lepton  around the former momentum, which is called
a normal term,
and another contribution from  it of broad spectrum around the latter
momentum, which is called a diffraction term.

The transition probability to this final state is  finite and
an order of  integrations are interchangeable. So the probability  is  
written, after the spin summations are made, with a  correlation
function and the neutrino wave function in the form 
 \begin{align}
&\int  \frac{d{\vec
 p}_l}{(2\pi)^3} \sum_{s_1,s_2}|T|^2 \nonumber\\
&=   \frac{N_2}{E_\nu}\int d^4x_1 d^4x_2 
e^{-\frac{1}{2\sigma_\nu}\sum_i ({\vec x}_i-\vec{x}_i^{\,0})^2} \Delta_{\pi,l}(\delta x)
e^{i \phi(\delta x)},
\label{probability-correlation} 
\end{align}
where $N_2=g^2
\left({4\pi}/{\sigma_{\nu}}\right)^{\frac{3}{2}}V^{-1}$, $V$ is
a normalization volume for the initial pion, $\vec{x}_i^{\,0} = \vec{\text{X}}_{\nu} + {\vec
v}_\nu(t_i-\text{T}_{\nu})$, $\delta x
=x_1-x_2$, $\phi(\delta x)=p_{\nu}\!\cdot\!\delta x $
  and 
\begin{align}
\Delta_{\pi,l} (\delta x)=&
 {\frac{1}{(2\pi)^3}}\int
{d {\vec p}_l \over E({\vec p}_l)}(2p_{\pi}\cdot p_{\nu} p_{\pi}\cdot
 p_l-m_{\pi}^2 p_l \cdot p_{\nu})\nonumber\\  
&\times  e^{-i(p_{\pi}-p_l)\cdot\delta x }. \label{pi-mucorrelation}
%& &\delta t=t^1-t^2,\delta {\vec x}={\vec x}^1-{\vec x}^2.\nonumber
\end{align}
In Eq.\,$(\ref{pi-mucorrelation} )$, the  momentum, $p_l$, is integrated
in whole positive energy  region.

\section{Light-cone singularity}
 
 $\Delta_{\pi,l} (\delta x)$ is composed of  a light-cone
  singularity \cite{Ishikawa-Tobita-prl1,Wilson-OPE} and 
  regular terms. 
The former is generated from those plane waves that have a 
same phase and are added constructively.
  In order to extract a leading singular term in the variable, $\delta x$,
  we  write  the integral  in a four dimensional form 
with a  new variable  $q=p_{l}-p_{\pi}$ that is conjugate to $\delta
 x$. Then  $\Delta_{\pi,l}(\delta x)$ is
 decomposed into   integrals  in $0 \leq q^0$ and   $-p_{\pi}^0 \leq
q^0 \leq 0$. 
An  integral from $0 \leq q^0$ is written in the form, $\left\{2(p_{\pi}\cdot p_{\nu}) p_{\pi}\cdot
 \left(p_{\pi}-i{\partial \over
 \partial \delta x}\right)-m_{\pi}^2\left(p_{\pi}-i{\partial \over
 \partial \delta x}\right)  \cdot p_{\nu}\right\}  \tilde I_1$, where 
\begin{align}
\tilde I_1=\int d^4 q \,  \frac{\theta(q^0)}{4\pi^4}\text {Im}\left[1 \over
 q^2+2p_{\pi}\!\cdot\! q+{\tilde m}^2-i\epsilon\right] e^{iq \cdot \delta x
 }\nonumber,
\end{align}
and  ${\tilde m}^2=m_{\pi}^2-m_{l}^2$. 
The  integrand of $\tilde I_1$ is expanded in
$p_{\pi}\!\cdot\! q$
and  the integration leads  the light-cone singularity \cite{Wilson-OPE}, $\delta({\delta
x}^2)$,  and less singular and regular terms  which are described with  
Bessel functions.  
An integral from the region  $-p_{\pi}^0 \leq
q^0 \leq 0$, $I_2$, is written  with the momentum $\tilde
q=q+p_{\pi}$ and has no singularity. 
Thus the correlation function, $\Delta_{\pi,l}(\delta x)$, 
  is written in the form 
\begin{align}
&\Delta_{\pi,l}(\delta x)=2i
\left\{2(p_{\pi}\cdot p_{\nu}) p_{\pi}\cdot
 \left(p_{\pi}-i\frac{\partial}{
 \partial \delta x}\right)-m_{\pi}^2\left(p_{\pi}-i{\partial \over
 \partial \delta x}\right)  \cdot p_{\nu}
\right\}\nonumber\\
&\times  \left[ D_{\tilde m}\left(-i\frac{\partial}{\partial \delta x}\right) 
\left( \frac{\epsilon(\delta t)}{4\pi}\delta(\lambda)  
+f_{short}\right) +I_2\right],\label{muon-correlation-total}
\end{align}
where $\lambda=\delta x^2$, $D_{\tilde m}(-i\frac{\partial}{\partial
\delta x})$
$=$
$\sum_l$
$ (1 /l!)\bigl(2p_{\pi}\!\cdot\!(-i{\partial \over \partial \delta x})$
$\frac{\partial}{\partial {\tilde m}^2}\bigr)^l$, $f_{short}$
$=$
$-\frac{i{\tilde m}^2}{
 8\pi \xi }$
$\theta(-{\lambda})$
$\{N_1(\xi
 )-$
$i$
$\epsilon(\delta t)$
$J_1(\xi)\}
 -$
$\frac{i{\tilde m }^2}{
 4\pi^2\xi }$
$\theta(\lambda)$
$K_1(\xi)$, $\xi=\tilde m\sqrt \lambda $, $N_1$, $J_1$, and $K_1$ are 
Bessel functions.
$f_{short}$ has a singularity  of the form $1/\lambda$ around
$\lambda =0$ and decrease as $e^{-\tilde m\sqrt{|\lambda}|}$ or
oscillates as  $e^{i\tilde m \sqrt{|\lambda}|}$ at a large
$|\lambda|$. 

The series in Eq.\,$(\ref{muon-correlation-total})$ converges when
$2p_{\pi}\!\cdot\!p_{\nu}\leq {\tilde m}^2$ and this expression is valid in
this kinematical region.

The singular terms, $\delta(\lambda)$ and others, in Eq.\,$(\ref{muon-correlation-total})$ decrease slowly with 
the distance and give a 
correlation effect in a wide area. These  terms are derived from the 
integration in $E_{\pi} \leq E_l$. Because
this region is outside of kinematical  region of satisfying the energy 
and momentum  conservation $p_{\pi}=p_{\nu}+p_l$, its contribution disappears 
in the amplitude defined at $\text{T}=\infty$. Conversely
  the singular terms are  included  only in the  physical quantities
  observed at a finite time
  interval. So they are not derived from  the standard calculations of plane
  waves with the asymptotic conditions. 
   The latter term, $I_2$, on the other hand, comes from the integration
   region 
$E_l \leq E_{\pi}$, which is in the kinematical region of 
satisfying the energy and momentum conservation. So
  this  determines the quantities  at $\text{T}=\infty$. This term oscillates or
  decreases fast in $\lambda$ with a time scale 
  determined with   ordinary microscopic quantities and becomes 
microscopically short, as most other cases. We will see that 
the light-cone singularity is combined with the small neutrino mass 
and gives a finite distance correction of an exceptional scale to 
the position dependent neutrino flux.

\section{Integration of space-time  coordinates} 

Next,
Eq.\,$(\ref{muon-correlation-total})$ is substituted to
Eq.\,$(\ref{probability-correlation})$ and  the coordinates ${\vec x}_1$ and ${\vec x}_2$ are integrated. 
The most singular term, $J_{\delta(\lambda)}$, is from $\frac{\epsilon(\delta t)}{4\pi}\delta(\lambda) $ in
$\Delta_{\pi,l}$, which has no scale,     
 and is rewritten using the center coordinate $X^\mu=({
x_1^\mu+x_2^\mu})/{2}$ and the relative coordinate
$\vec{r}=\vec{x}_1-\vec{x}_2$.
The center coordinate $\vec{X}$ is integrated  easily and  $J_{\delta(\lambda)}$ becomes
an integral of  the relative coordinates $({\vec
r}_T,r_l)$.   
Finally we have 
\begin{align}
J_{\delta(\lambda)}=C_{\delta(\lambda)}
{\epsilon(\delta t)}{|\delta t|^{-1}
 }e^{i\bar \phi_c(\delta t)-\frac{m_{\nu}^4}{
 16\sigma_{\nu} E_{\nu}^4} {\delta
 t}^2},\label{lightcone-integration2-2}
\end{align}
where $C_{\delta(\lambda)}={(\sigma_{\nu}\pi)}^{\frac{3}{2}}
 {\sigma_{\nu}}/{2}$ and $\bar\phi_c(\delta t)=\omega_{\nu} \delta t,\omega_{\nu}={m_{\nu}^2 / 2E_{\nu}}$. The phase $\phi(\delta x)$ of
 Eq.\,$(\ref{probability-correlation})$ becomes  the small phase 
$\bar \phi_c(\delta t)$ of Eq.\,$(\ref{lightcone-integration2-2})$ at the
 light cone due to a cancellation between the time and space
 components.The angular velocity of  $\bar{\phi}_c(\delta t)$ is 
extremely small and  the interference is not 
 determined with de Broglie phase but  this small phase. This leads
an  unusual neutrino interference of the present  work. 
The next singular 
term is from  
${1/\lambda}$ in $\Delta_{\pi,l}$, and becomes  $ J_{\delta(\lambda)} / \sqrt {\pi \sigma_{\nu}
 |\vec{p}_{\nu}|^2}$, which is much smaller than $ J_{\delta(\lambda)}$ in the
 present parameter region.
 The
 magnitude is inversely proportional to  $|\delta t|$ and is 
independent from 
 the ${\tilde m}^2$ for the general form of wave packet also.

  Integrating   
$t_1$ and $t_2$ in a finite $\text{T}=\text{T}_{\nu}-\text{T}_{\pi}$, 
we have  a slowly decreasing    term    $\tilde g(\text{T},\omega_{\nu})$
 and a normal term $G_0$.  $\tilde
g(\text{T},\omega_{\nu})$ is generated from the light-cone singularity 
and related
term.
$\tilde g(\text{T},\omega_{\nu})=g(\text{T},\omega_{\nu})-g(\infty ,\omega_{\nu})$ where  
\begin{eqnarray}
& & \text{T}  g(\text{T},\omega_{\nu})=-i  \int_0^{\text{T}} dt_1 dt_2  \frac{\epsilon(\delta t)}{|\delta t|}e^{i {\omega_{\nu}}\delta t }, \label{probability1} 
\end{eqnarray}
and  $\tilde g(\infty,\omega_{\nu})=0$.
The normal term,    $\text{T} G_0$, is from the rest.
 Due to the rapid oscillation in $\delta t$, $G_0$  gets
 contribution  from 
 the microscopic  $\delta t$ region and  is constant in a macroscopic T. This 
term does not depend on $\sigma_\nu$ and agrees with  the normal probability 
obtained with the standard method of using plane waves. 
In the region $2 p_{\pi}\!\cdot\!p_{\nu} >
\tilde{m}$,  $\Delta_{\pi,l} (\delta x)$ does not have the
light-cone singularity and the diffraction term  exists  only in 
the kinematical region 
$2 p_{\pi}\!\cdot\!p_{\nu}\leq \tilde{m}$.  

  We compute  a total 
probability next. From an integration of neutrino's coordinates
$\vec{\text{X}}_{\nu}$,  the total volume $V$ is obtained and  cancelled with 
the normalization of the initial pion state. The total
probability, then, becomes a sum of the normal term $G_0$ and the
diffraction term $\tilde g(\text{T},\omega_\nu)$, 
\begin{align}
\label{probability-1}
P=N_3\int \frac{d^3 p_{\nu}}{(2\pi)^3}
\frac{p_{\pi}\! \cdot\! p_{\nu}(m_{\pi}^2-2p_{\pi}\! \cdot\! p_{\nu}) }{E_\nu}
 \left[\tilde g(\text{T},\omega_{\nu}) 
 +G_0 \right],
%\label{probability-1}, 
\end{align}
where $N_3 = 8\text{T}g^2 \sigma_\nu$ and $\text{L} = c\text{T}$ is the
length of a decay region. 
At $\text{T} \rightarrow \infty$, the diffraction term vanishes and the
probability $P$ agrees with the value of the standard
calculation of plane waves. At a finite T, the probability has the
diffraction component, $\tilde g(\text{T},\omega_{\nu})$, which is stable under 
  variation of the pion's energy.        
%%%%%%%%%%%%%%%%%%%%%%%%%%%%%%% FIG of angle and neutrino energy %%%%%%%%%%%%%%%%%%%
 \begin{figure}[t]%
  \centering{
   {\includegraphics[angle=-90,scale=.3]{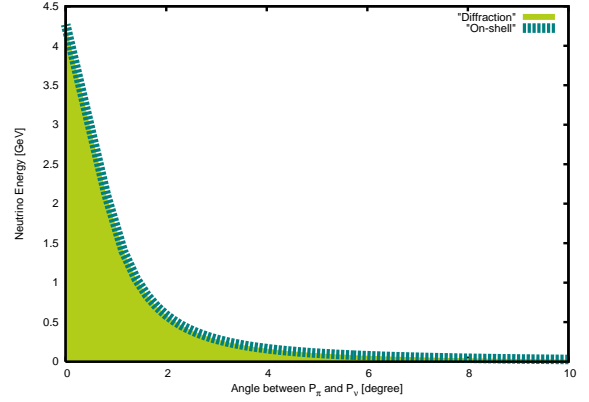}}%	
  }
  \caption{The angle between $p_{\pi}$ and $p_{\nu}$
  dependence of the neutrino energy is   given.   The horizontal axis 
  shows the angle and the vertical axis shows the neutrino energy at
  $E_\pi = 10$\,[GeV]. The normal term has a value along the boundary and the
  diffraction term has a value in a broad area below the normal term.}
  \label{fig:angle-energy}
 \end{figure}%
%%%%%%%%%%%%%%%%%%%%%%%%%%%%%%% FIG of angle and neutrino energy %%%%%%%%%%%%%%%%%%%

Now we study  each term of Eq.\,$(\ref{probability-1})$. In the normal 
term, $G_0$, the energy and momentum are  conserved well, and  $G_0$ has a sharp peak  at 
$p_{\pi}\!\cdot\! p_{\nu}={\tilde{m}^2/2}$. Hence the factor
 $m_{\pi}^2-2p_{\pi}\cdot p_{\nu}$ in Eq.\,$(\ref{probability-1})$ becomes
 $m_l^2$ and a cosine of angle between ${\vec p}_{\pi}$ and ${\vec
 p}_{\nu}$ is determined uniquely by the relation,
\begin{align}
\cos \theta_{\pi,\nu}=\frac{(m_l^2-m_{\pi}^2+2E_{\pi}E_{\nu})}{ 2|\vec{p}_{\pi} ||\vec{p}_{\nu}|},
\label{angle-normal}
\end{align}
and the probability  is proportional to $m_l^2$. Integration of  
the neutrino's angle leads
  this
  integral  independent of the angle width, as far as it
  include the narrow peak. The value   is independent also  of 
$\sigma_{\nu}$, which is consistent
 with  the condition for the stationary state \cite{Stodolsky}.

The diffraction component, $\tilde g(\text{T},\omega_{\nu})$, on the 
other hand, is present in a kinematical region,  
$|\vec{p}_{\nu}|(E_{\pi}-|\vec{p}_{\pi}|)\leq p_{\pi}\!\cdot\! p_{\nu}
\leq {\tilde{m}^2/2}$.
The energy and momentum are not exactly conserved in the space-time
dependent amplitude Eq.\,$(\ref{amplitude})$.  Hence 
$m_{\pi}^2-2p_{\pi}\!\cdot\! p_{\nu}$ in 
Eq.\,$(\ref{probability-1})$ is larger than  $m_l^2$, and  the cosine of  angle between ${\vec p}_{\pi}$ and ${\vec
 p}_{\nu}$   is not  uniquely determined.
The dependence of the diffraction term upon the angle between
$\vec{p}_{\pi}$ and  $\vec{p}_{\nu}$ is 
 presented  in 
Fig.\,\ref{fig:angle-energy} 
for
${ E_\pi = 10}$\,[GeV]. The angle, which is determined uniquely 
from the neutrino energy for
the small wave packet, is not unique and is widely spread.  
From this behavior of  the diffraction
   component, it might have been   possible to reject the diffraction  
component and to take only the normal component in narrow band beam 
experiments if  the  neutrino's interaction position  were  observed.  
Comparison of the
   total cross section of satisfying this constraint with that of 
non-constraint   events might have given direct signals of the diffraction component.

 The diffraction term is slowly varying with both the distance and energy.
The typical length $\text{L}_0$ of this universal behavior  is  
$\text{L}_0~[\text{m}] ={2E_{\nu} \hbar c / m_{\nu}^2 }= 400\times {E_{\nu}[\text{GeV}]/
 m_{\nu}^2[\text{eV}^2/c^4]}$. 
Its magnitude  was found to be  about  10-20 \%\ of the 
normal term in a suitable  experimental situation and depends
on geometry of the detector. In this paper we analyze   total  neutrino 
nucleon cross section, which is connected with the neutrino flux and its high energy behavior is understood
well from a standard quark parton model. 

\section{ Charged leptons}

A probability   of observing  a neutrino or a charged lepton  in the 
pion decay at a position of a macroscopic distance  is written 
 in  the form \cite{Ishikawa-Tobita-prl1,Ishikawa-Tobita-prl2},
\begin{eqnarray}
P= P_{normal}+P_{diff}^{l}.
\label{probability-lepton}
\end{eqnarray}
In Eq.\,$(\ref{probability-lepton})$,  $P_{normal}$ is  the normal
term  that is obtained  from the decay probability $G_0$ in
Eq.\,$(\ref{probability-1})$ and $P_{diff}^l$ is  the diffraction  term that
is determined from $\tilde g$ in Eq.\,$(\ref{probability-1})$. 
The latter has not been  included
in calculating   the total cross sections before and its effect is
estimated here.

\begin{figure}[t]
\includegraphics[scale=.3,angle=-90]{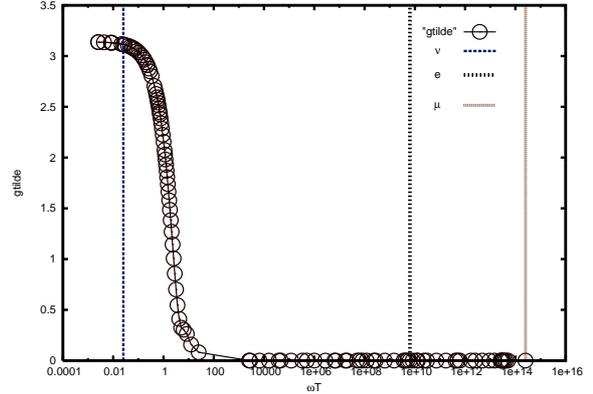}
\caption{Values of $\tilde g(\text{T},\omega)$ for $\nu$, $e$ and $\mu$
 at $\text{L}=10$\,[m], $E=1$\,[GeV] and $m_{\nu}c^2=1$\,[eV], $m_{e}c^2=0.5$\,[MeV]
 and $m_{\mu}c^2=100$\,[MeV]. }
\label{fig:g}
\end{figure}

The diffraction term  at a time $\text{T}$ is described by its mass
 and energy   
in the universal form
\begin{eqnarray}
P_{diff}^l=C_{diff}\text{T} \tilde g(\text{T},\omega_l),
\label{diffraction}
\end{eqnarray}
where $C_{diff}$ is a constant and is obtained later.   $\omega_l$ are small in neutrinos
and large in charged leptons. 
$\tilde g(\text{T},\omega_l)$ is positive definite and decreases slowly with a 
distance $\text{L}=c\text{T}$, and  vanishes at the infinite  distance with the 
length scale  ${2cE_l /m_l^2}$. This scale is  macroscopic  
for   neutrinos  but  is $10^{-10}$\,[m] or less for the electron and
muon. The magnitude of
 $\tilde g(\text{T},\omega_l)$ is given in Fig.\,($\ref{fig:g}$). At $\text{L}=10\,[\text{m}]$, 
$E=1\,[\text{GeV}]$ for the mass
 $1\,[\text{eV}/{c^2}]\,(\nu)$, $0.5\,[\text{MeV}/{c^2}]\,(e)$ and $100\,[\text{MeV}/{c^2}]\,(\mu)$, the
 values are, 
\begin{align}
&\tilde g(\text{T},\omega_{\nu} ) \approx 3, \nonumber\\
&\tilde g(T,\omega_e ) \approx 0, \nonumber \nonumber\\
&\tilde g(T,\omega_{\mu} ) \approx 0.
\end{align}
%%%%%%%%%%%%%%%% figure of \tilde{g} %%%%%%%%%%%%
%%%%%%%%%%%%%%%% figure of \tilde{g} %%%%%%%%%%%%%
Hence the diffraction component at a macroscopic distance is finite in 
neutrinos and vanishes in others. 
It
is striking that  the  neutrino flux has an additional term and 
 is not equivalent to  that of the charged lepton even though they are
 produced in the same decay process.  This diffraction term
     is generated by  the tiny neutrino mass and 
the  interference term generated with the superposed waves
of forming the light-cone singularity. Because  $P_{diff}$   is the interference term, it has  
unusual  properties  different from those of $P_{normal}$ in  the
neutrino flavour 
and the  energy and momentum. The neutrino diffraction furthermore is  
sensitive to the absolute neutrino mass.   We study implications of the
diffraction term to  total cross sections, next.

\section{Total cross sections of $\nu_{\mu}$-N scattering}

A total cross section of a neutrino nucleon scattering  is  written in
the form,
\begin{eqnarray}
\label{total-crossection}
& &\sigma^{\nu}={M_N E_{\nu}G_F^2 \over \pi }(Q+\bar Q/3),
\end{eqnarray}
using integrals of quark-parton distribution functions $q(x)$ and $\bar
q(x)$,  $Q=\int_0^1 dx xq(x),\bar Q=\int_0^1 dx x\bar q(x)$. The cross
sectin  is 
proportional to the neutrino energy and a current value is
$\sigma_{\nu}/E= (0.677\pm0.014) \times
10^{-38}\,[\text{cm}^2/\text{GeV}]$ \cite{particle-data}. So experiments seem consistent with 
Eq.\,$(\ref{total-crossection})$.  However  recent experiments of
NOMAD \cite{excess-near-detectorNOMAD} and MINOS \cite{excess-near-detectorMino} gave the total cross sections in wide energy ranges with
small uncertainties and showed that the cross sections have slight
energy dependences. They  are compared with our theoretical calculations
in the following.

The diffraction term was identified only recently.
Here  we include the diffraction term  into the neutrino flux. 
 The neutrino flux in pion decays is given as a sum of the normal and
diffraction terms in the form 
\begin{eqnarray}
f=f_{normal}(1+r_{diff}),
\end{eqnarray}
where $r_{diff}$ is a ratio  of the diffraction component over the
normal component, and is a function of   $\zeta =
 {m_{\nu}^2L/2cE_{\nu}}
$,
\begin{eqnarray}
r_{diff}=d_{0} \tilde g(\zeta),
\label{E-depedent-correction}
\end{eqnarray}
where  the coefficient $d_0$
is determined from  geometry. 

When the detector is
located at an end of a decay volume, the correction
factor Eq.\,$(\ref{E-depedent-correction})$ is used. In an actual
experiment, the
detector is located in a distant region from a decay volume. There is 
soil between them and pions are stopped in a beam dump. The neutrino
diffraction does not occur and neutrino propagates freely in this
region. Since the wave packets of one $\sigma_\nu$ form a complete
set, the wave packet size at the decay volume is the $\sigma_\nu$
determined with the detector.
The 
neutrino flux at the end of the decay volume is computed with the
diffraction term of the decay volume 's length $\text{L}$ and the wave packet
size of the detector. 
Wave packets of this $\sigma_\nu$  propagate
 freely from the end of a decay volume to the detector. The final value of 
the neutrino flux  at the detector is found using the factor
Eq.\,$(\ref{E-depedent-correction})$. When a neutrino changes flavour  in
this period, the final probability for each flavour is written with  a
standard  formula of flavour oscillation. 
 
A true value of the total cross section, $\sigma(E)^{true}$, obtained
with the total flux is connected
with a  cross section, $\sigma(E)^{exp}$, obtained with only the normal 
component of the  flux
by  the ratio   
\begin{eqnarray}
\sigma(E)^{true}
=\sigma(E)^{exp} { 1 \over1+ r_{diff} }.\nonumber 
\end{eqnarray}
Conversely the experimental   cross section is written as 
\begin{eqnarray}
\sigma(E)^{exp}/E= ( 1+ r_{diff} )(\sigma(E)^{true}/E).
\label{energy-dependece}
\end{eqnarray}
$\sigma(E)^{true}/E$ is believed a constant so  the E-dependence of 
$\sigma(E)^{exp}/E$ is due to the E-dependence of $ r_{diff}$,
Eq.\,$(\ref{E-depedent-correction})$.  

Since the diffraction term has different properties  from those of 
the normal term, the corrections $ r_{diff}$ depends  also on the geometry of 
the experiment and the material of the detector. In this paper 
we compute the  cross sections using the experimental conditions 
of  MINOS  and MOMAD, which have presented precise cross sections of 
various energy ranges under the same condition, and compare  them with the
experiments.

%%%%%MINOS%%%%%%%%%%%
\begin{figure}[t]
\includegraphics[scale=.3,angle=-90]{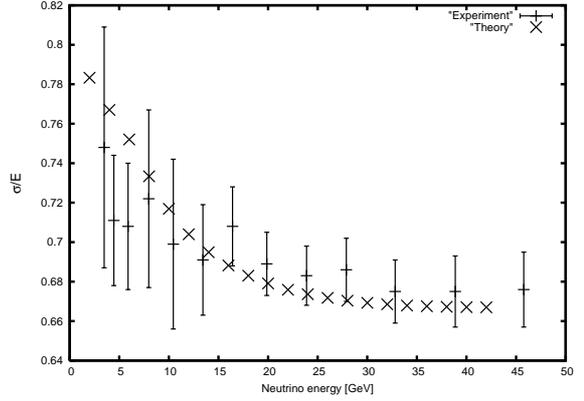}
\caption{Total cross section of MINOS is compared with the sum of normal
 and diffraction terms.$m_{\nu}c^2=0.2eV$}
\label{MINOS:fig}
\end{figure}
%%%%%%MINOS%%%%%%%%%%%%%

%%%%%NOMAD%%%%%%%%%%%
\begin{figure}[t]
\includegraphics[scale=.3,angle=-90]{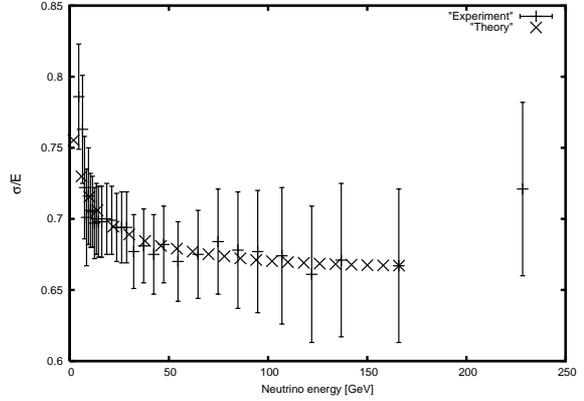}
\caption{Total cross section of NOMAD is compared with the sum of normal
 and diffraction terms.$m_{\nu}c^2=0.2eV$}
\label{NOMAD:fig}
\end{figure}
%%%%%%NOMAD%%%%%%%%%%%%%

The geometry of MINOS and  NOMAD are the following. The lengths between the
pion source and the neutrino detector, $L_{det-so}$, and those  
of the decay regions, $\text{L}_{dec-reg}$, 
 are:
\begin{align}
\text{NOMAD}~:&~\text{L}_{det-so}=835\,[\text{m}], ~\text{L}_{dec-reg}=290\,[\text{m}],\nonumber \\
\text{MINOS}~:&~\text{L}_{det-so}=1040[\text{m}], ~\text{L}_{dec-reg}=675\,[\text{m}]. \nonumber 
\end{align} 
The detector size is  $7\times3\times3 \,[\text{m}^3]$ for both experiments.

The wave packet size is estimated  using the size of target nucleus. 
From the size of the nucleus 
of the mass number $A$, we have  $\sigma_{\nu}= A^{\frac{2}{3}}/m_{\pi}^2$. 
For various material   the value are
\begin{align}
&  \sigma_{\nu}= 5.2/m_{\pi}^2;~ {}^{12}C~ nucleus,\nonumber \\
& \sigma_{\nu}= 14.3/m_{\pi}^2;~ {}^{54}Fe ~nucleus.
\end{align}

The total cross sections of MINOS and NOMAD experiments are compared with 
the theoretical values
in Figs.\,$(\ref{MINOS:fig})$ and $(\ref{NOMAD:fig})$. In theoretical
calculations, effects of a pion beam spreading is included by taking 
an average of initial pion's angle  from $0$ to $10\,[\text{mrad}]$ for NOMAD and
from $0$ to $15\,[\text{mrad}]$ for MINOS.  An effect due to a finite 
size of the initial pion was also estimated. We found that an plane wave
approximation which was employed in this paper  was very good.

The cross 
sections
decrease quite slowly with the neutrino energy, which may be difficult to
understand  with the standard theory. The theoretical cross sections obtained 
by including the diffraction component into the neutrino flux 
showed the same behavior and agreed well with the experiments. Two
experiments  are actually different in the neutrino energy and geometry,
but  agreed with the theory.
So the large cross sections at low energy regions may be attributed to 
   the diffraction component. 

We have compared only NOMAD and MINOS here. Many experiments are listed 
in particle data \cite{particle-data} and most of them have similar
energy dependences and agree qualitatively with the diffraction's
presence. It is important to notice 
that the magnitude of diffraction component is sensitive to geometry
and if a kinematical constraint Eq.\,($\ref{angle-normal}$) on the angle 
between ${\vec
p}_{\pi}$ and  ${\vec p}_{\nu}$ was required, only the events of
the normal term was selected. Then  the cross section should agree with
that of the normal term.

\section{Summary and implications}

We showed that due to  the diffraction component, $P_{diff}$,  the
neutrino flux was  modified. 
The total neutrino cross sections of NOMAD and MINOS  agreed  
with the theoretical calculations obtained using   the
modified neutrino flux. Thus the existence of  the
neutrino diffraction is consistent with experimental observations of the 
total cross sections.  Although the diffraction component is determined
by the average neutrino mass, its sensitivity of the existing data is not
sufficient to find an information on the neutrino mass below $0.3\,[\text{eV}/c^2]$.

Other channels of interests are quasi-elastic or one pion production
processes. The cross sections for  
$\nu+n \rightarrow \mu^{-}+p^+\ (+\pi^0),\nu+p^+ \rightarrow \mu^{-}+p^++\pi^{+}$
 and $\bar \nu+p^+ \rightarrow \mu^{+}+n\ (+\pi^0)$ and the neutral current
 process  are known well theoretically using models such as CVC, PCAC, and 
vector dominances and others. Recent experiments \cite{excess-MiniBoone}
showed that the cross sections have excesses of $20-50$\%\ and are consistent
with the diffraction terms.   Especially a proton is known to have large wave
packet size in matter due to its small mass and the diffraction due 
to a proton target is enhanced and gives a finite contribution despite its
small energy ratio, $m_{proton}/m_{nucleus}$.

{\bf Acknowledgements.} One of the authors (K.I) thanks Drs. Kobayashi,
Nishikawa, Maruyama and Nakaya  for useful discussions on 
the neutrino detectors, Drs. Asai, Kobayashi, Minowa, Mori and Yamada
for useful discussions on interferences. 

{}

\end{document}